\documentclass[11pt]{article}
\usepackage{fullpage}
\usepackage{color}
\usepackage{graphicx}

\newcommand{\ceiling}[1]{\lceil #1\rceil}
\newcommand{\floor}[1]{\lfloor #1\rfloor}

\newcommand{\rankhigh}{\textsf{rank\_high}}
\newcommand{\ranklow}{\textsf{rank\_low}}

\newtheorem{observation}{Observation}
\newtheorem{theorem}{Theorem}

\title{Simplified, stable parallel merging}
\author{Jesper Larsson Tr\"aff\\
Faculty of Informatics, Institute of Information Systems, 
Research Group Parallel Computing\\
Vienna University of Technology (TU Wien)\\
Favoritensta\ss e 16, 1040 Vienna\\
Austria
}

\begin{document}
\maketitle

\begin{abstract}
This note makes an observation that significantly simplifies a number
of previous parallel, two-way merge algorithms based on binary search
and sequential merge in parallel.  First, it is shown that the
additional merge step of distinguished elements as found in previous
algorithms is not necessary, thus simplifying the implementation and
reducing constant factors. Second, by fixating the requirements to the
binary search, the merge algorithm becomes stable, provided that the
sequential merge subroutine is stable. The stable, parallel merge
algorithm can easily be used to implement a stable, parallel merge
sort.

For ordered sequences with $n$ and $m$ elements, $m\leq n$, the
simplified merge algorithm runs in $O(n/p+\log n)$ operations using
$p$ processing elements. It can be implemented on an EREW PRAM, but
since it requires only a single synchronization step, it is also a
candidate for implementation on other parallel, shared-memory
computers.
\end{abstract}


\section{Introduction}

Many parallel, two-way merge algorithms (see, e.g.,
\cite{Gavril75,GerbessiotisSiniolakis01,HagerupRub89,KatajainenLevcopoulosPetersson93,Kruskal83,ShiloachVishkin81})
for the case where the number of elements in the sequences to be
merged is larger than the number of processing elements build
on the scheme found in, e.g.~\cite{ShiloachVishkin81}: Binary search
is used to divide the two input sequences into disjoint sequences that
can be merged pairwise independently. Binary searches are performed in
parallel for a small selection of \emph{distinguished elements} from
the two input sequences. A separate, parallel merge of distinguished
and located elements is needed to determine pairs of subsequences to
merge. Often, such algorithms are not naturally stable, and take extra
space to be made stable. This note shows that the parallel merge of
distinguished and located elements is not needed, and at the same time
makes the merge algorithm stable without any additional space or time
overhead. This significantly simplifies implementation, whether on a
PRAM~\cite{JaJa92,Traff00:prambook} or on real hardware with, e.g.,
OpenMP~\cite{OpenMP2001}. On an EREW PRAM the simplified algorithm
runs in $O(n/p+\log n)$ time steps where $n$ is the size of the
longest input sequence.

A different approach to two-way merging than the one sketched above
was introduced in~\cite{AklSantoro87}, and used subsequently
in~\cite{DeoJainMedidi94,DeoSarkar91}
and~\cite{VarmanIyerHaderleDunn90,VarmanScheuflerIyerRicard91}. With
this approach the blocks of the input sequences that are needed to
produce any given block of the output sequence are identified directly
by a special procedure similar to binary search. This has the
advantage of giving the processing elements the same number of input
elements to merge, which is achieved only to within a factor of two by
the above approach. The observation in this note is not relevant to
this class of parallel merge algorithms.

\section{The merge algorithm}

Let $A$ and $B$ be two non-decreasing sequences ordered by a relation
$\leq$ with $n$ and $m$ elements, respectively. Both sequences are
allowed to contain repeated elements, and are also not required to be
distinct. Assume without loss of generality that $m\leq n$. The input
sequences are stored in arrays indexed from $0$, and a merged output
sequence is to be delivered in an array $C$ with $n+m$ elements. For
convenience, assume that $A[-1]=-\infty,A[n]=\infty$, and similarly
for array $B$. An implementation does not have to store or reserve
space for these sentinel elements, though.

\begin{figure}
\begin{flushleft}
\begin{tabular}{r|llll|llll|llll|lll|lll|l}
& $x_0$ & & & & $x_1$ & & & & $x_2$ & & & & $x_3$ & & & $x_4$ & & & $x_5$ \\
$i$: & 0 & 1 & 2 & 3 & 4 & 5 & 6 & 7 & 8 & 9 & 10 & 11 & 12 & 13 & 14 & 15 & 16 & 17 & \\
\cline{2-19}
$A[i]$: & 0 & 0 & 1 & 1 & 1 & 2 & 2 & 2 & 4 & 5 & 5 & 5 & 5 & 5 & 6 & 6 & 7 & 7 & \\
\cline{2-19}
& & & & & & $\bar{y_0}$  & & & $\bar{y_1}$ & $\bar{y_2}$ & & & & & & & $\bar{y_3}$ & & $\bar{y}_4$ \\
& & & & & & & & & & & & & & & & & & & $\bar{y}_5$ \\
\end{tabular}
\\[\bigskipamount]
\end{flushleft}
\begin{flushleft}
\begin{tabular}{r|lll|lll|lll|lll|lll|l}
& $\bar{x}_0$ & \\
& $\bar{x}_1$ & & & & & & $\bar{x}_2$ & $\bar{x}_3$ & $\bar{x}_4$ & & & & & & & $\bar{x}_5$ \\
$j$: & 0 & 1 & 2 & 3 & 4 & 5 & 6 & 7 & 8 & 9 & 10 & 11 & 12 & 13 & 14 & 15 \\
\cline{2-16}
$B[j]$: & 1 & 1 & 3 & 3 & 3 & 3 & 4 & 5 & 6 & 6 & 6 & 6 & 7 & 7 & 7 & \\
\cline{2-16}
& $y_0$ & & & $y_1$ & & & $y_2$ & & & $y_3$ & & & $y_4$ & & & $y_5$ \\
\end{tabular}
\end{flushleft}
\caption{Two non-decreasing sequences $A$ and $B$ with $n=18$ and
  $m=15$ elements, respectively, divided into $p=5$ consecutive
  blocks. For the starting elements $x_i$ and $y_j$ of the blocks the
  corresponding low and high \emph{cross ranks} are shown, denoted as
  $\bar{x}_i$ and $\bar{y}_j$, respectively.  The cross ranks from the
  $A$ array illustrate four of the five cases for the merge step:
  $x_0$ (a), $x_1$ and $x_2$ (e), $x_3$ (b), and $x_4$ (c). The cross
  ranks $\bar{y}_0$ and $\bar{y}_3$ from $B$ illustrate case (d).  The
  algorithm identifies the following $2p=10$ merge subproblems of
  disjoint sequences that can be handled in parallel: $A[0,\ldots,3]$
  is copied into $C[0,\ldots,3]$, $A[4]$ is copied into $C[4]$, $A[8]$
  is copied into $C[14]$, $A[12,\ldots,14]$ and $B[7]$ are merged into
  $C[19,\ldots,22]$, $A[15]$ and $B[8]$ are merged into $C[23,24]$
  (all by Step~\ref{step:ABmerge}); and $B[0,\ldots,2]$ and
  $A[5,\ldots,7]$ are merged into $C[5,\ldots,10]$, $B[3,\ldots,5]$ is
  copied into $C[11,\ldots,13]$, $B[6]$ and $A[9,\ldots,11]$ are
  merged into $C[15,\ldots,18]$, $B[9,\ldots,11]$ and $A[16,17]$ are
  merged into $C[25,\ldots,29]$, $B[12,\ldots,14]$ is copied into
  $C[30,\ldots,32]$ (all by Step~\ref{step:BAmerge}).}
\label{fig:crossrank}
\end{figure}

For some element $x$ and array $X$ let its \emph{low rank}, denoted
$\ranklow(x,X)$, be the unique index $i, 0\leq i\leq n$ such that
\begin{displaymath}
X[i-1]<x\leq X[i]
\end{displaymath}
and its \emph{high rank}, denoted $\rankhigh(x,X)$, be the unique
index $j$ such that
\begin{displaymath}
X[j-1]\leq x<X[j]
\end{displaymath}

The low (resp.\ high) rank of an element $a=A[i]$ (resp.\ $b=B[i]$) in
the array $B$ (resp.\ $A$) will be referred to as the \emph{cross
  rank} of $a$ (resp.\ $b$) in $B$ (resp.\ $A$). The correctness of
the merge algorithm is based on the observation that cross ranks ``do
not cross''. By this, cross ranks from $A$ can be used to partition
$B$ and vice versa. Consider the cross ranks of two elements $A[x_i]$ and
$A[x_{i+1}]$ with $x_i<x_{i+1}$ in $B$. The observation below states
  that the cross rank of any element in $B$ between the cross ranks of
  $A[x_i]$ and $A[x_{i+1}]-1$ will be between $x_i$ and $x_{i+1}$.
Cross ranks for selected elements of the two sequences are
shown in Figure~\ref{fig:crossrank}.

\begin{observation}
Let $a=A[i]$ be an element of $A$ and let $j=\ranklow(a,B)$ be its cross
rank in $B$. The cross rank for any element $j'<j$ that comes before
$j$ in $B$ is not after $i$, i.e., $\rankhigh(B[j'],A)\leq i$, and the
cross rank for any element $j''>j$ that comes after $j$ in $B$ is
strictly after $i$, i.e., $\rankhigh(B[j''],A)>i$. In particular, for
$i'=\rankhigh(B[j],A)$ it is the case that $i'>i$. The same holds
\emph{mutatis mutandis} for elements in $B$.
\end{observation}

To see this, consider the cross rank $j=\ranklow(a,B)$ of an element
$a=A[i]$, and let $j'<j$. Since $j$ is the low rank of $a$ in $B$,
$B[j']\leq B[j-1]<a\leq B[j]$, so therefore $\rankhigh(B[j'],A)\leq
i$. For $j<j''$, $a\leq B[j]\leq B[j'']$ and therefore
$\rankhigh([B[j''],A)>i$.

Both low and high ranks can be computed by suitably modified binary
search in $O(\log n)$ steps. The low rank of $a=A[i]$ from $A$ is the
number of elements from $B$ that must come before $a$ in a stable
merge of $A$ and $B$ in which all (repeated) elements $a$ from $A$ are
ordered before elements $a$ from $B$, that is the position of $a$ in
the stably merged output sequence is $i+\ranklow(A[i],B)$. Conversely,
the high rank of $b=B[j]$ in $A$ is the number of elements from $A$
that must come before $b$ in a stable merge, that is the position of
$b$ in the stably merged output is $j+\rankhigh(B[j],A)$.

Now, let $p$ be the number of processing elements. The input sequences
$A$ and $B$ are divided into roughly equal sized, consecutive,
contiguous \emph{blocks} differing in size by at most one. The first $r=n\bmod
p$ blocks of $A$ will get $\ceiling{n/p}$ elements, and the remaining
blocks $\floor{n/p}$ elements; similarly for the $B$ array.  The start
index $x_i$ of block $i$ in $A$ for each $0\leq i<p$ is determined by

\begin{eqnarray*}
x_i & = & \left\{
\begin{array}{ll}
i\ceiling{n/p} & \mbox{for}\ i<r \\
i\floor{n/p}+n\bmod p & \mbox{otherwise}
\end{array}
\right.
\end{eqnarray*}

\begin{figure}
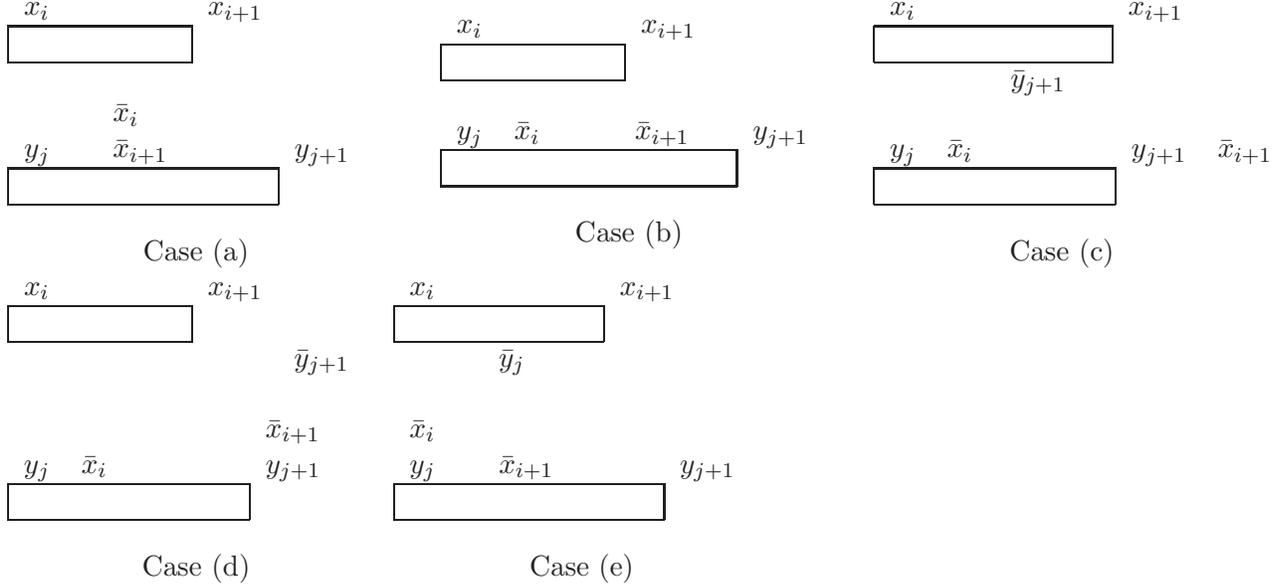

\begin{minipage}{5cm}
\begin{flushleft}
\begin{tabular}{|lllll|l}
\multicolumn{1}{l}{$x_i$} & & & & \multicolumn{1}{l}{} & $x_{i+1}$ \\
\cline{1-5}
& & & & & \\
\cline{1-5}
\end{tabular}
\\[\bigskipamount]
\begin{tabular}{|llllll|l}
\multicolumn{1}{l}{} & & $\bar{x}_i$ & & & \multicolumn{1}{l}{} & \\
\multicolumn{1}{l}{$y_j$} & & $\bar{x}_{i+1}$ & & & \multicolumn{1}{l}{} & $y_{j+1}$\\
\cline{1-6}
& & & & & & \\
\cline{1-6}
\end{tabular}
\end{flushleft}
\centering{Case (a)}
\end{minipage}
\begin{minipage}{5cm}
\begin{flushleft}
\begin{tabular}{|lllll|l}
\multicolumn{1}{l}{$x_i$} & & & & \multicolumn{1}{l}{} & $x_{i+1}$ \\
\cline{1-5}
& & & & & \\
\cline{1-5}
\end{tabular}
\\[\bigskipamount]
\begin{tabular}{|llllll|l}
\multicolumn{1}{l}{$y_j$} & $\bar{x}_i$ & & & $\bar{x}_{i+1}$ & \multicolumn{1}{l}{} & $y_{j+1}$\\
\cline{1-6}
& & & & & & \\
\cline{1-6}
\end{tabular}
\end{flushleft}
\centering{Case (b)}
\end{minipage}
\begin{minipage}{5cm}
\begin{flushleft}
\begin{tabular}{|lllll|l}
\multicolumn{1}{l}{$x_i$} & & & & \multicolumn{1}{l}{} & $x_{i+1}$ \\
\cline{1-5}
& & & & & \\
\cline{1-5}
\multicolumn{1}{l}{} & & & $\bar{y}_{j+1}$ & \multicolumn{1}{l}{} & \\
\end{tabular}
\\[\bigskipamount]
\begin{tabular}{|llllll|ll}
\multicolumn{1}{l}{$y_j$} & $\bar{x}_i$ & & & & \multicolumn{1}{l}{} &
$y_{j+1}$ & $\bar{x}_{i+1}$ \\
\cline{1-6}
& & & & & & \\
\cline{1-6}
\end{tabular}
\end{flushleft}
\centering{Case (c)}
\end{minipage}
\begin{minipage}{5cm}
\begin{flushleft}
\begin{tabular}{|lllll|ll}
\multicolumn{1}{l}{$x_i$} & & & & \multicolumn{1}{l}{} & $x_{i+1}$ & \\
\cline{1-5}
& & & & & & \\
\cline{1-5}
\multicolumn{1}{l}{} & & & & \multicolumn{1}{l}{} & & $\bar{y}_{j+1}$ \\
\end{tabular}
\\[\bigskipamount]
\begin{tabular}{|llllll|l}
\multicolumn{1}{l}{} & & & & & \multicolumn{1}{l}{} &
$\bar{x}_{i+1}$ \\
\multicolumn{1}{l}{$y_j$} & $\bar{x}_i$ & & & & \multicolumn{1}{l}{} &
$y_{j+1}$ \\
\cline{1-6}
& & & & & & \\
\cline{1-6}
\end{tabular}
\end{flushleft}
\centering{Case (d)}
\end{minipage}
\begin{minipage}{5cm}
\begin{flushleft}
\begin{tabular}{|lllll|l}
\multicolumn{1}{l}{$x_i$} & & & & \multicolumn{1}{l}{} & $x_{i+1}$ \\
\cline{1-5}
& & & & & \\
\cline{1-5}
\multicolumn{1}{l}{} & & $\bar{y}_{j}$ & & \multicolumn{1}{l}{} & \\
\end{tabular}
\\[\bigskipamount]
\begin{tabular}{|llllll|l}
\multicolumn{1}{l}{$\bar{x}_i$} & & & & & \multicolumn{1}{l}{} & \\
\multicolumn{1}{l}{$y_j$} & & $\bar{x}_{i+1}$ & & & \multicolumn{1}{l}{} & $y_{j+1}$\\
\cline{1-6}
& & & & & & \\
\cline{1-6}
\end{tabular}
\end{flushleft}
\centering{Case (e)}
\end{minipage}
\caption{The five cases identifying the possible subsequences to merge
  for the processing element assigned to start index $x_i$ of $A$
  based on the cross ranks from $x_i$ and $x_{i+1}$. Cross ranks from
  start indices $y_i$ and $y_{i+1}$ of $B$ similarly identifies the
  subsequences to merge when processing elements are assigned to
  $y_i$.}
\label{fig:fivecases}
\end{figure}

In addition, define $x_p=n$. The start indices $y_i$ for the $p$
blocks of $B$ are defined similarly. Let $k$ be some index in $A$
(resp.\ $B$). Index $k$ belongs to block $i$ if either
$k<r\ceiling{n/p}$ and $\floor{k/\ceiling{n/p}}=i$, or $k\geq
r\ceiling{n/p}$ and $\floor{(k-r\ceiling{n/p})/\floor{n/p}}+r=i$.
Computing a block start index $x_i$ or $y_i$ as well as determining
the block to which a given index $k$ belongs are thus all constant
time operations. With this, stable, parallel merge is accomplished by
the following steps:

\begin{enumerate}
\item
Compute $\bar{x}_i=\ranklow(A[x_i],B)$ for $0\leq
i<p$ by binary search in parallel. Also, let $\bar{x}_p=m$.
\item
Compute $\bar{y}_j=\rankhigh(B[y_j],A)$ for $0\leq
 j<p$ by binary search in parallel. Also, let $\bar{y}_p=n$.
\item
\label{step:ABmerge}
Merge disjoint sequences from $A$ and $B$ in parallel by assigning a
processing element to each $x_i$ for $0\leq i<p$ as follows:
\begin{enumerate}
\item
If $\bar{x}_i=\bar{x}_{i+1}$ output $A[x_i,\ldots,x_{i+1}-1]$ to 
$C[x_i+\bar{x}_i,\ldots]$.
\item
If $\bar{x}_i\neq\bar{x}_{i+1}$ are in the same block $j$ of $B$, and
$\bar{x}_i\neq y_j$, merge $A[x_i,\ldots,x_{i+1}-1]$ stably with 
$B[\bar{x}_i,\ldots,\bar{x}_{i+1}-1]$ and output to $C[x_i+\bar{x}_i,\ldots]$.
\item
If $\bar{x}_i$ and $\bar{x}_{i+1}$ are in different $B$ blocks, with
$\bar{x}_i$ in block $j$, $\bar{x}_i\neq y_j$ and $\bar{x}_{i+1}\neq y_{j+1}$, 
merge $A[x_i,\ldots,\bar{y}_{j+1}-1]$ stably with
$B[\bar{x}_i,\ldots,y_{j+1}-1]$. Output to $C[x_i+\bar{x}_i,\ldots]$.
\item
If $\bar{x}_i$ and $\bar{x}_{i+1}$ are in different $B$ blocks, with
$\bar{x}_i$ in block $j$, $\bar{x}_i\neq y_j$ and $\bar{x}_{i+1}=y_{j+1}$, 
merge $A[x_i,\ldots,x_{j+1}-1]$ stably with
$B[\bar{x}_i,\ldots,y_{j+1}-1]$. Output to $C[x_i+\bar{x}_i,\ldots]$.
\item
If $\bar{x}_i$ is in block $j$ of $B$, $\bar{x}_i = y_j$, and
$\bar{x}_i\neq\bar{x}_{i+1}$, output $A[x_i,\ldots,\bar{y}_j-1]$ to
$C[x_i+\bar{x}_i,\ldots]$.
\end{enumerate}
\item
\label{step:BAmerge}
Merge disjoint sequences from $B$ and $A$ in the same fashion by
assigning a processing element to each $y_j$ for $0\leq j<p$.
\end{enumerate}

The results of the binary searches are shown in
Figure~\ref{fig:crossrank} which illustrates in particular the five
cases for the merge steps. By the observation on cross ranks, all
blocks and sequences are disjoint, and obviously partition the
arrays. That the five cases are exhaustive is illustrated in
Figure~\ref{fig:fivecases}. In particular, for case (c), the condition
implies that $\bar{x}_{i+1}\geq y_{j+1}$ and therefore it holds that
$\bar{y}_j\leq x_{i+1}$, such that the segment
$A[x_i,\bar{y}_{j+1}-1]$ falls entirely within block $i$ of $A$. The
exception where $\bar{x}_{i+1}=y_{j+1}$ is covered by case (d); by the
observation it namely holds that $\bar{y}_{j+1}>x_{i+1}$.  Correctness
is therefore established.  Stability in the sense that all elements
$a$ of $A$ will be placed before elements $a$ of $B$ is also
maintained by the use of low and high ranks, provided a stable
sequential merge is used. Since all blocks contain $O(n/p)$ elements
and the sequences determined by the cross ranks as in the cases (a) to
(e) fall entirely within blocks, the number of steps for the parallel
merge operations with $p$ processing elements is likewise
$O(n/p)$. The binary searches are done in parallel and take $O(\log
n)$ operations. Determining which merge case applies entails
determining the blocks of $\bar{x}_i$ and $\bar{x}_{i+1}$ and takes
$O(1)$ operations. In summary:

\begin{theorem}
Two ordered sequences $A$ and $B$ of length $n$ and $m$, respectively,
with $m\leq n$ can be merged stably in parallel in $O(n/p+\log n)$
operations using $p$ processing elements. Only constant extra space in 
addition to the input and output arrays is needed.
\end{theorem}

Synchronization is only required after the two binary search steps,
before which the cross ranks $\bar{x}_i$ and $\bar{y}_j$ are
conveniently stored in $p+1$ element arrays. The algorithm can
trivially be implemented on a CREW PRAM. For implementation on an EREW
PRAM, it is first observed that each merge of a block from $A$
requires comparing only two locations, namely $\bar{x}_i$ and
$\bar{x}_{i+1}$, so accessing indices $\bar{x}_i$ and $\bar{x}_{i+1}$
by the $p$ processing elements in different steps suffices to
eliminate concurrent reads. Start addresses of the arrays $A$, $B$,
and $C$ can be copied to the $p$ processing elements in $O(\log p)$
steps by parallel prefix operations. The parallel binary searches can
be pipelined to eliminate concurrent reads, and still the $p$ searches
can be done in $O(\log n)$ parallel time steps. Also this is a
standard technique~\cite{AklMeijer90}. A more processor efficient
algorithm for EREW PRAM parallel binary search can be found
in~\cite{Chen95:pbs}. Although the sizes of the blocks to merge are
all $O(n/p)$, the sizes of the blocks that are merged by different
processing elements can differ by a factor of two. Cases (a) and (e)
can give rise to blocks with a small, constant number of elements,
wheres as the cases (b)-(d) can give rise to blocks with almost
$2\floor{n/p}$ elements.


Overall, the algorithm is a considerable simplification
of~\cite{HagerupRub89,ShiloachVishkin81} and other similar merging
algorithms in that no merging of the sequences of distinguished
elements is needed.

\section{Applications and remarks}

The stable parallel merge algorithm can be used to implement a stable,
parallel merge sort that runs in $O(n\log n/p+\log p\log n)$ parallel
time steps for $n$ element arrays. As in~\cite{ShiloachVishkin81} this
is done by first sorting sequentially in parallel $p$ consecutive
blocks of $O(n/p)$ elements, and then merging the sorted blocks in
parallel in $\ceiling{\log p}$ merge rounds. In round $i$,
$i=1,...,\ceiling{\log p}$ there are at most $\ceiling{p/2^i}$ pairs
to be merged and possibly one sequence that just has to be
copied. This can be accomplished either by grouping the processing
elements into groups of $2^i$ consecutively numbered elements, or by
modifying the merge algorithm to work in parallel on the
$\ceiling{p/2^i}$ pairs. The latter can easily be accomplished. Thus,
a stable merge sort can be implemented with no extra space apart from
input and output arrays.

The simplified merge algorithm is likewise useful for distributed
implementation, e.g. on a BSP as in~\cite{GerbessiotisSiniolakis01};
here the eliminated merge of $p$ pairs of distinguished elements can
save at least one expensive round of communication. Details are
outside the scope of this note.

\bibliographystyle{abbrv}
\bibliography{parallel,traff}

\end{document}